\documentclass[]{elsart}
\usepackage{amssymb}
\usepackage{pst-node,pstricks,graphicx}
\begin{document}                % INITIALIZE - DONT CHANGE
%%\flushright{Draft}
\begin{frontmatter}
\title{Variation of atmospheric depth profile on different time scales}
\author[a]{B.~Wilczy\'nska\corauthref{cor1}},
\author[a]{D.~G\'ora},
\author[a]{P.~Homola},
\author[a]{J.~P\c{e}kala},
\author[a,b]{M.~Risse},
\author[a]{and H.~Wilczy\'nski}
\corauth[cor1]{ {\it Correspondence to}: B.~Wilczy\'nska
(Barbara.Wilczynska@ifj.edu.pl)}
\address[a]{
Institute of Nuclear Physics PAN, 
ul.Radzikowskiego 152, 
31-342 Krak\'ow, Poland
}
\address[b]{
Forschungszentrum Karlsruhe, Institut f\"ur Kernphysik, 76021 Karlsruhe, Germany
}
\begin{abstract}

The vertical profile of atmospheric depth is an important element in extensive air shower studies. The depth of shower maximum is one of the most important characteristics of the shower. In the fluorescence technique of shower detection, the geometrical reconstruction provides the altitude of shower maximum, so that an accurate profile of atmospheric depth is needed to convert this altitude to the depth of shower maximum. In this paper the temporal variation of experimentally measured profiles of atmospheric depth at different sites is studied and  implications for shower reconstruction are shown. The atmospheric profiles vary on time scales from hours to years. It is shown that the daily variation of the profile is as important as its seasonal variation and should be accounted for in air shower studies. For precise shower reconstruction, the daily profiles determined locally at the site of the air shower detector are recommended.

\end{abstract}


\end{frontmatter}
%%%%%%%%%%%%%%%%%%%%%%%%%%%%%%%%%%%%%%%%%%%%%%%%%%%%%%%%%%%%%%%

\section{Introduction}

The atmosphere is the key component of any detection system for extensive air showers. All cosmic rays with energies larger than about $10^{15}$ eV are detected via recording 
extensive air showers they initiate in the atmosphere. Details of shower development 
depend on the primary cosmic ray particle, but also on the properties of the 
atmosphere itself, in particular on the vertical distribution of mass in the atmosphere. Since the 
atmosphere serves both as a target in which primary cosmic rays interact and as the medium in which showers develop, a 
precise knowledge of relevant properties of the atmosphere is extremely important for 
studies of high energy cosmic rays. 

Studies of the atmosphere done over decades have resulted in a fairly good knowledge 
of the distributions of atmospheric air. The atmospheric conditions are quantitatively 
described using parameters like temperature, pressure, wind speed etc. as a function 
of time and location.  In contrast to many other areas of physics in 
which measurements and observations are performed in laboratory experiments and are well 
controlled and reproducible, the atmospheric phenomena are studied by observing 
natural processes which cannot be controlled by the observer. These atmospheric 
phenomena are reproducible only as much as is characteristic to the phenomena under 
study. 
For example, the temporal variation of temperature can be presented as a superposition of 
an average over many years, seasonal variations, diurnal variation and irregular 
variation of different time scales. Similarly one can use the spatial scale (vertical or 
horizontal) of  the temperature field or of other parameters of 
interest.

The atmosphere is in a state very near the local hydrostatic equilibrium, i.e. the difference of 
atmospheric pressures between two levels is always close to the weight of a column of air 
with unit cross section. This results in a characteristic, near exponential, vertical distribution of pressure (and 
consequently of air mass) in the atmosphere. Generally, the atmospheric pressure (and the overlying 
mass) falls by an order of magnitude every 16--20 km, for altitudes up to 100 km. 

Traditionally, the atmosphere is divided into layers, most commonly according to the 
vertical distribution of temperature. These layers are:
{\em (i)} troposphere, in which temperature falls with increasing altitude, which extends up to  
altitudes of about 7 km near the poles or 16 km near the equator,   
{\em (ii)} stratosphere, up to about 50 km with approximately constant temperature (lower 
stratosphere) or increasing temperature (upper stratosphere),
{\em (iii)} mesosphere, up to about 85 km, in which temperature falls again with altitude,
{\em (iv)} thermosphere,  where the temperature increases with altitude, and which extends up 
to the limit of atmosphere.
The boundaries between these layers are called tropopause, stratopause and 
mesopause, respectively.
Most of the mass of the atmosphere (about 75\%) is contained in the troposphere; the troposphere  and 
stratosphere combined contain more than 99\% of the atmospheric mass.

The primary source of energy which drives the atmospheric phenomena is the radiation 
of the Sun. The local flux of incoming energy depends on geographic location, shape of 
the terrain, time of day and year and on transparency of the air (i.e. cloudiness, pollution 
etc). Movements of air make the dependence on irradiation more complex, therefore observations and measurements of atmospheric phenomena must be done in 
many places to get a coherent picture of the atmosphere. 

The main atmospheric parameter which governs the longitudinal development of an extensive air shower is the column depth of the air traversed by the shower over its pathlength $l$ ("slant depth"): $ X_{sl} = \int \rho(h) dl $, where $\rho(h)$ is the altitude-dependent air density. Therefore, for studies of air showers, the basic atmospheric parameter is the vertical profile of atmospheric depth: $ X_V(h) = \int_h^\infty \rho(h) dh $, which provides a correspondence between the atmospheric depth $X_V$ and altitude $h$ above the sea level. For showers with not-too-large zenith angles, $\theta \le 70^\circ$, the vertical and slant atmospheric depth are related by $X_{sl} = X_V / \cos\theta$. In the following, mainly the vertical atmospheric depth will be considered, so for simplicity $X$ will be used instead of $X_V$, unless explicitly stated otherwise.

In studies related to development and detection of extensive air showers, the US Standard Atmosphere Model \cite{USstd} is commonly used. This model provides distributions of temperature and pressure as functions of altitude. The 1976 extension of the Model provides also the northern mid-latitude winter and summer atmosphere distributions. 
An important question is, however, how well does the US Standard Atmosphere Model approximate local conditions at the sites of air shower detectors, and what is the time variability of the local atmosphere? In other words, is the annual or seasonal average adequate for a particular day at a particular location? This question is especially important for the fluorescence technique of shower detection. The depth of shower maximum, $X_{max}$, is the key parameter for identification of the primary particle of the shower. However, in the fluorescence detector the altitude $h_{max}$ of shower maximum is determined from the geometry reconstruction and the corresponding atmospheric depth of the maximum has to be derived using an appropriate $X \leftrightarrow h$ relation.
It is now well known \cite{Keilh-AP22,BW-Tsuk,BK-Tsuk,JB-Pune,BW-Pune} that a good knowledge of atmospheric depth profile is essential for precise shower reconstruction.

The global COSPAR International Reference Atmosphere (CIRA86) \cite{CIRA} provides temperature and pressure profiles at altitudes above 20 km at many latitudes at both hemispheres. However, most of the air shower development takes place at altitudes smaller than 20 km, so that the CIRA86 model is not sufficient for air shower studies.

The UK Met Office, through the British Atmospheric Data Centre,  maintains a data base of measurements done by radiosondes 
at a number of locations worldwide \cite{BADC}. A radiosonde is a small package of instruments, suspended from a helium-filled 
balloon. During ascent of the balloon the radiosonde measures the temperature, pressure 
and humidity at intervals of typically 2 seconds. The information is transmitted via radio 
to the ground station, which processes and stores the data. The balloon typically reaches 
the altitude 20--30 km before it bursts and the radiosonde falls, being attached to a small parachute. 
The "standard resolution" data, which are used in this study, consist of radiosonde data 
read out at the standard pressure levels, which are: 1000, 925, 850, 700, 500, 400, 300, 
250, 200, 150, 100, 70, 50, 30, 20 and 10 mbar. 

In this paper, we present the results of a study of atmospheric profiles at Salt Lake City (USA), near the site of the HiRes experiment and at Mendoza (Argentina), near the site of the southern Pierre Auger Observatory. The studies are based on the Met Office data collected in daily balloon soundings of the atmosphere performed at the Salt Lake City and Mendoza airports.

\section{Variation of temperature and pressure profiles}

\begin{figure}
\hspace{-0.5cm}
\includegraphics*[width=1.1\textwidth,angle=0,clip]{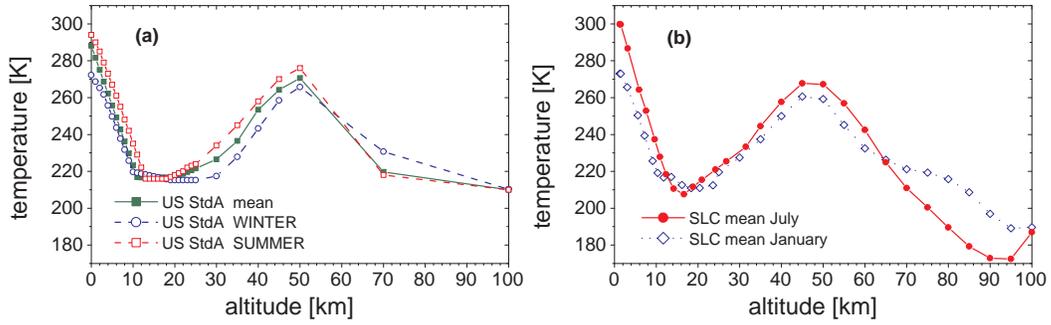}
\caption {Temperature dependence on altitude in {\em (a)} the US Standard Model and {\em (b)} averages of measurements done in January and July at Salt Lake City. }
\label{USmodel}
\end{figure}

The variation of temperature with altitude in the US Standard Atmosphere Model is shown in Fig. \ref{USmodel} for annual mean as well as for winter and summer. The variation of actual measurements done at Salt Lake City in winter (average of January measurements over four years) and in summer (average of July) are also shown. The seasonal variation of temperature distribution is apparent. Although a similar behaviour of the temperature distributions in both the model and the measurements is observed, there are also considerable differences, which will be discussed below.

\begin{figure}
\begin{center}
\hspace{-10mm}
\includegraphics*[width=0.55\textwidth,angle=0,clip]{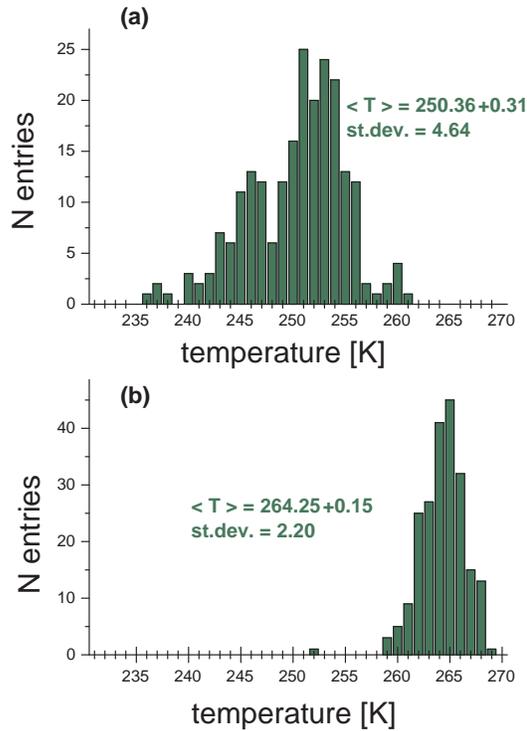}
\end{center}
\caption{Temperature distribution in {\em (a)} January and {\em (b)} July at altitude 5.8 km at Salt Lake City. }
\label{temperatures}
\end{figure}

In the following, let's concentrate on fluctuations of the air temperature. Since the temperature of the air near the ground is strongly influenced by weather effects, choosing some higher altitude seems to be more representative for studies of the seasonal variation of the temperature distribution. We choose the altitude of 5.8 km above sea level (near the 500 mbar standard pressure level in the Met Office data set).  The distribution of daily temperature at this altitude during the months of January and July is shown in Fig. \ref{temperatures}. These come from soundings made at Salt Lake City during years 1999--2002. The average temperatures in January and July at 5.8 km altitude differ by 14 degrees. The variation of temperature in January is much larger than in July. The width of the distribution, especially for the January data, suggests that in addition to seasonal differences, the daily variation may also be important.

\begin{figure}
\begin{center}
\hspace*{-10mm}
\includegraphics*[width=0.9\textwidth,angle=0,clip]{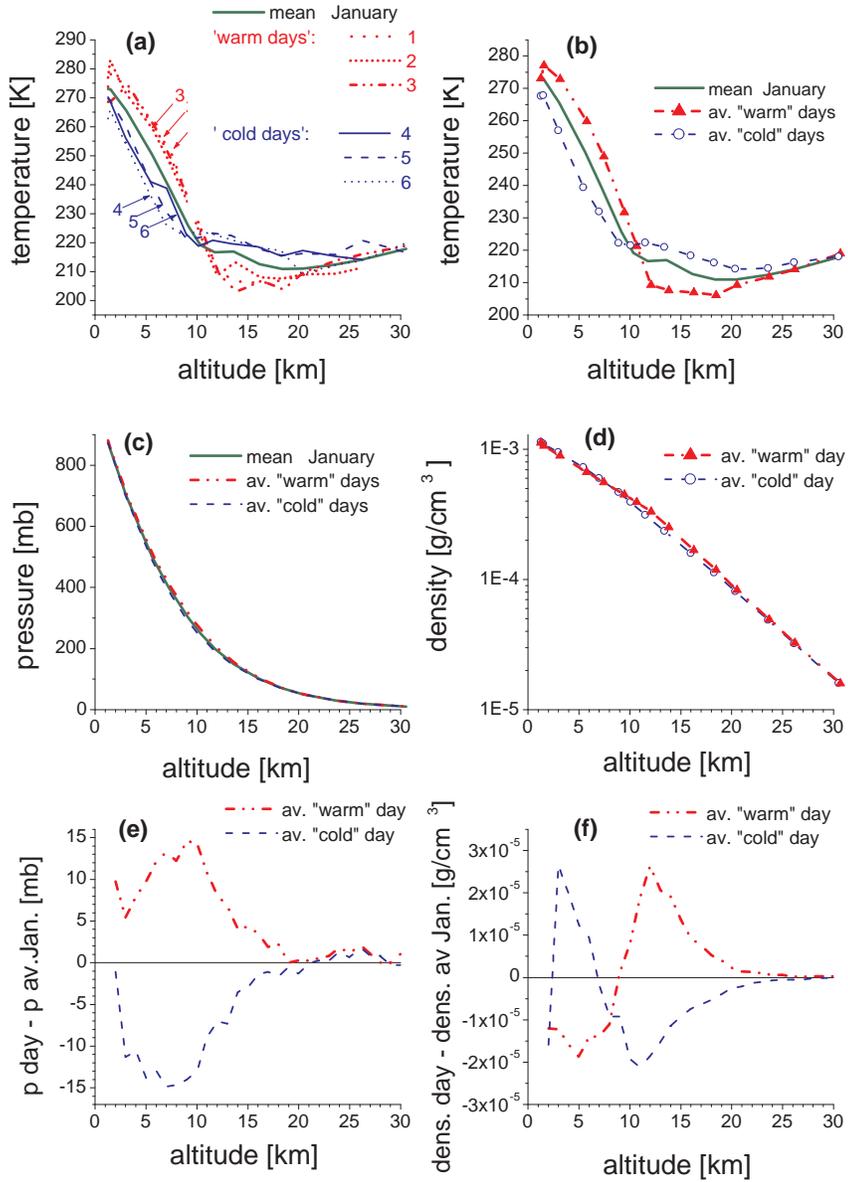}
\end{center}
\caption{{\em (a, b)} Temperature  and {\em (c)} pressure distributions in three extremely warm and  three extremely cold days in January at Salt Lake City. {\em (d)} Average air density in warm and cold days. {\em (e)} Differences in pressure distributions and {\em (f)} differences in density distributions of warm and cold days relative to the monthly averages.}
\label{warmcold}
\end{figure}

From the January Salt Lake City measurements, three extremely warm days (about 10 degrees warmer than average) were selected, to be compared with three cold days, with temperatures about 10 degrees lower than average. The temperature profiles of these selected days, shown in Fig. \ref{warmcold}a and \ref{warmcold}b, are quite distinct: a day warmer than average at altitude 5.8 km is colder than average in the tropopause (above $\sim$12 km) and vice versa. A "warm" day is consistently warmer than average over the whole troposphere and a "cold" day is consistently colder. The temperatures fall monotonically with altitude in the troposphere, with roughly the same rate of temperature decrease. We note that temperature inversions over small ranges of altitude are not visible in this standard resolution data set of measurements at relatively large intervals in altitude.
The profiles of pressure as a function of altitude are shown in Fig. \ref{warmcold}c and \ref{warmcold}e for the same days. They also show systematic differences between warm and cold days. As a result of the temperature and pressure differences, the derived densities of air (see Section 3) differ significantly, as shown in Fig. \ref{warmcold}d and \ref{warmcold}f. As the atmospheric depth is the integral of density of the overlying air, it is clear that the profiles of the atmospheric depth will be distinctly different for cold and warm days. 

\begin{figure}
\begin{center}
\hspace*{-10mm}
\includegraphics*[width=1.15\textwidth,angle=0,clip]{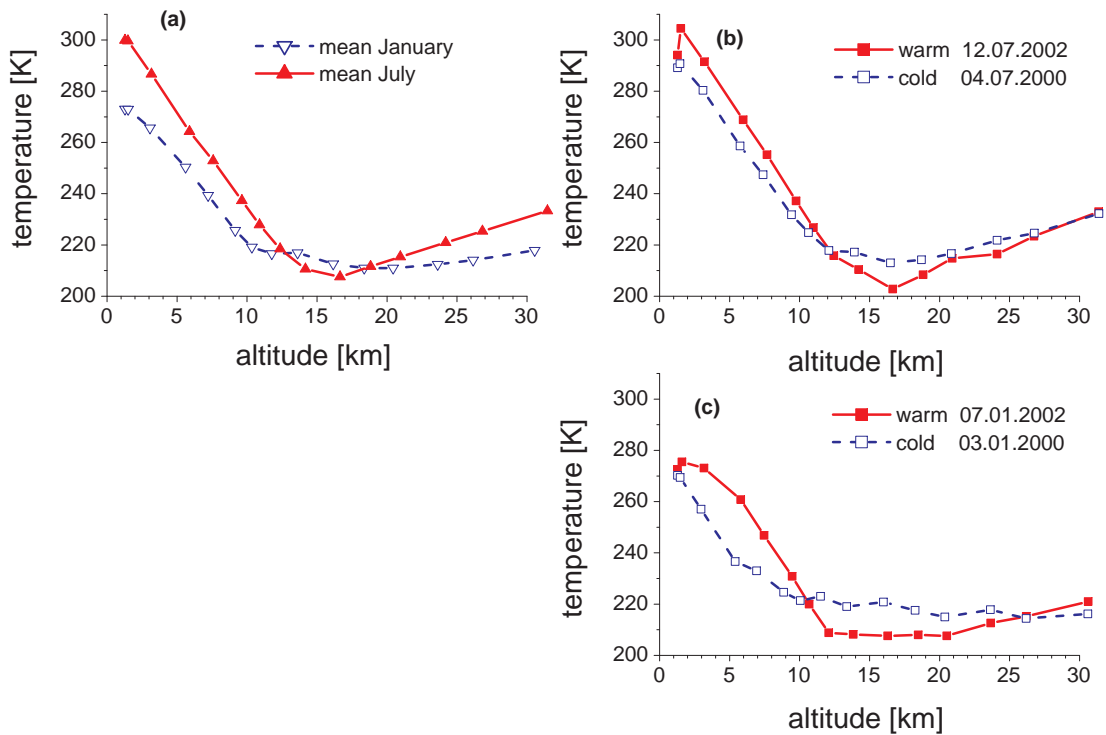}
\end{center}
\caption{Comparison of temperature distribution in cold/warm seasons to cold/warm days at Salt Lake City.}
\label{warmcoldSLC}
\end{figure}

As shown in Fig. \ref{warmcoldSLC}, a similar pattern of temperature variation holds for cold and warm {\em seasons} (January and July) and for cold and warm {\em days} in any of these two seasons. Higher temperatures at the low altitudes imply lower temperatures in the tropopause. The same pattern is observed in Mendoza.

The inference one can make is that the daily variation of the atmosphere may be as important as the seasonal variation. If so, the seasonal and daily fluctuations of temperature and pressure may make the actual atmospheric depth profiles differ considerably from the average one, as will be shown below.

\section{Determination of density and atmospheric depth profiles} 

The atmospheric depth at an altitude $h$ is the integral of density of the
overlying air: $X(h) = \int_h^\infty \rho (h) dh$. Since the air density is
not measured directly, it must be inferred from the ideal gas law based on
measurements of pressure $p$ and temperature $T$: 
\begin{equation} 
\rho(p, T) = \frac{p M_{mol}}{RT} = 0.34839 \frac{p {\mathrm [mbar]}}{T {\mathrm [K]}} \left[\frac{{\mathrm g}}{{\mathrm cm}^3}\right],
\end{equation} where $M_{mol} = 28.966$ g/mol is the molar mass of air and $R = 8.31432$ J/(K mol) is the universal gas
constant. In this work we use the pressure and temperature profiles  measured by radiosondes up to the altitude of 20--30 km (as available). 
At higher altitudes, the CIRA86 model is used.

The atmospheric depth in the US Standard Atmosphere model in the CORSIKA shower simulation package \cite{CORSIKA} is parameterized in four altitude ranges below 100 km by exponential functions  
\begin{equation}X_i = a_i + b_i e^{-h/c_i}
\end{equation} and by a linear function 
\begin{equation}X=a_5 - b_5 h/c_5 \end{equation}
for altitudes 100--112.8 km. This parameterization is widely used in shower simulation and reconstruction codes.

\begin{figure}
\begin{center}
\hspace*{-10mm}
\includegraphics*[width=1.15\textwidth,angle=0,clip]{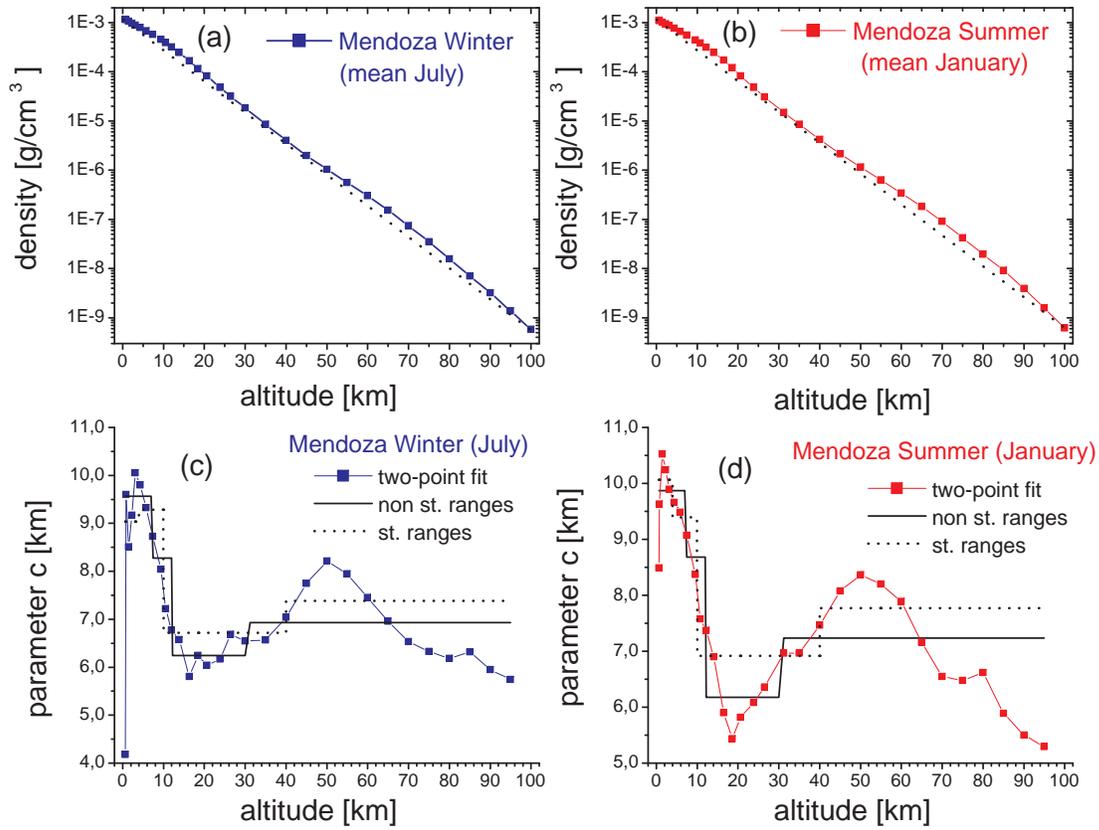}
\end{center}
\caption{Distribution of air density in {\em (a)} winter and {\em (b)} summer at Mendoza. In panels {\em (c)} and {\em (d)} the scale height parameter $c$, determined from the two-point fits (Eq.4) is compared to the four-exponent fits (Eq.2) in the standard altitude ranges (dotted lines) and in the chosen non-standard ranges (solid lines).}
\label{densityMza}
\end{figure}

The exponential form of the atmospheric depth profile implies an exponential distribution of the air density $\rho (h)$. The average July and January 
distributions of density in Mendoza are shown in Fig. \ref{densityMza}a and \ref{densityMza}b (the distributions at Salt Lake City are very similar). 
The exponential distributions should be represented by straight lines in these semi-logarithmic plots. The data points show the actual average air density in summer and in winter. The dotted lines are just straight lines drawn between the first and last data points in each plot. It is evident that the data points do not follow a single exponential distribution, nor exponential distributions are seen in specific ranges of altitude (no straight sections are present in the plots). 
A clear inference is that the real atmospheric density distribution does not quite follow the exponential distribution in any wide range of altitudes.  

To study the shape of the density distribution in more detail, the exponential function of the form 
\begin{equation}
\rho(h) = b \exp(-h/c)
\end{equation}
was fitted to every pair of adjacent data points. The variation of the parameter $c$ (the scale height of the exponential distribution) is shown in Figures \ref{densityMza}c and \ref{densityMza}d. This parameter changes continuously with altitude, so fitting the exponential function in wide ranges of altitudes is only a more or less crude approximation of the true distribution. Therefore, the procedure for determining the atmospheric depth profile $X(h)$ should be the following.
Instead of fitting the $\rho (h)$ function to the data points and integrating the $\rho(h)$ analytically to find $X(h)$, the atmospheric depth is determined by interpolating (with an exponential function of Eq. 4) the density between two adjacent data points and integrating the density from one data point to another. In this way, the obtained $\rho (h)$ distribution follows exactly the data points, and the uncertainties are due only to experimental errors of the data points, which are small. Assuming the measurement errors of temperature and pressure as quoted by the manufacturer of the radiosondes used, the resulting uncertainty of $X(h)$ turns out to be smaller than 2 g/cm$^2$ in the whole range of altitudes.

It is, however, convenient to use an analytical function to approximate the profile of air density (and consequently, atmospheric depth), as is commonly used (Eq. 2). The scale heights of the fits in the standard CORSIKA altitude ranges: 0--4 km, 4--10 km, 10--40 km and 40--100 km are shown in Fig. \ref{densityMza}c and \ref{densityMza}d by the dotted lines. It is evident that fits in other ranges can be more accurate. We choose fits in ranges: 0--7 km, 7--12 km, 12--30 km and 30--100 km, shown by the solid lines in \ref{densityMza}c and \ref{densityMza}d. Such "non-standard" ranges are allowed e.g. in CORSIKA. We note that the rather poor agreement of the fit with the data in the fourth range (above 30 km) has a negligible effect on air shower studies: the atmospheric depth at 30 km is only about 12 g/cm$^2$, so that showers develop mostly at lower altitudes.

\section{Seasonal variation of atmospheric profiles}

\begin{figure}[t]
\begin{center}
\hspace*{-10mm}
\includegraphics*[width=1.15\textwidth,angle=0,clip]{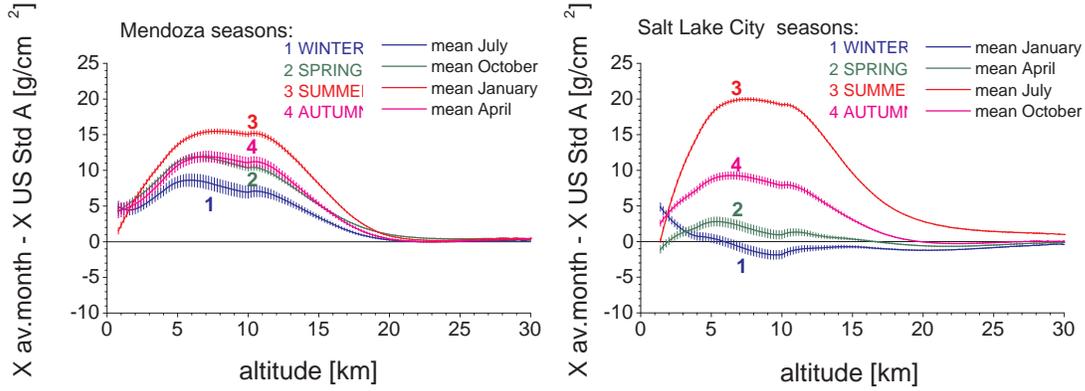}
\caption{\label {seas} Average seasonal atmospheric depth profiles, relative to the US Standard Atmosphere, at Mendoza (Argentina) and Salt Lake City (USA). The error bars represent the uncertainties of the average profiles.}
\end{center}
\end{figure}

\begin{figure}
\begin{center}
\hspace*{-10mm}
\includegraphics*[width=1.15\textwidth,angle=0,clip]{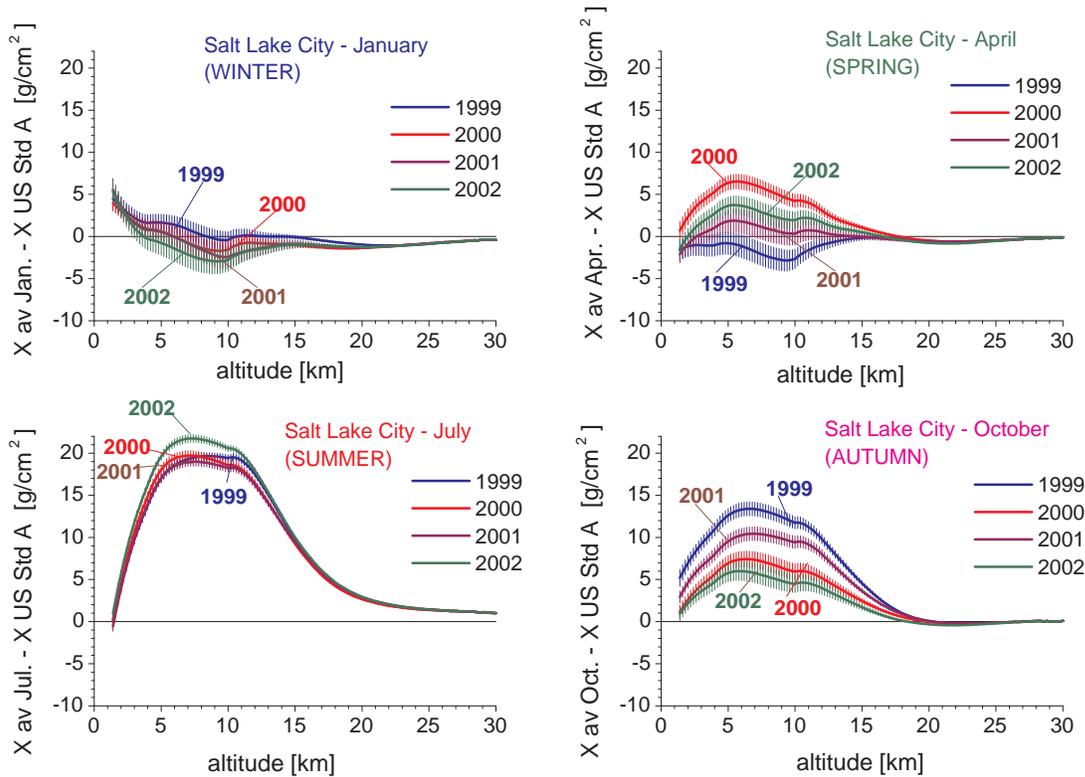}
\caption{\label {years} Year-to-year variability of monthly average profiles at Salt Lake City. The error bars represent the uncertainties of the average profiles.}
\end{center}
\end{figure}

Given the large seasonal variation of the temperature profile discussed in Section 2, it is to be expected that the profile of atmospheric depth may also vary considerably. The seasonal variation of the atmospheric depth profile $X(h)$ at both sites analyzed (Salt Lake City and Mendoza) is shown in Fig. \ref{seas}. To better show the details, the differences between the determined profiles and the US Standard Atmosphere model are plotted. Average profiles of single months chosen as representatives of each season are presented. A considerable seasonal variation is manifest, especially at Salt Lake City: the January and July profiles differ from each other by up to 20 g/cm$^2$ at Salt Lake City and by up to 10 g/cm$^2$ in Mendoza. At both sites the summer profile differs most from the US Standard Atmosphere model, and the winter profile is closest to it. The summer-winter variation is considerably larger at Salt Lake City than at Mendoza. It is to be noted that the US Standard Atmosphere Model is not adequate even for the annual average and would not represent an average local profile at either site.
The seasonal profiles shown in Fig. \ref{seas} are averages of four years for the month indicated. However, the monthly averages vary year-to-year, as shown in Fig. \ref{years}. A monthly average determined in one year can differ from that determined in the next year by up to 10 g/cm$^2$. This year-to-year variation of the averages is a consequence of the daily variation, as discussed below.

\section{Day-to-day variation}

\begin{figure}
\begin{center}
\hspace*{-10mm}
\includegraphics*[width=1.05\textwidth,angle=0,clip]{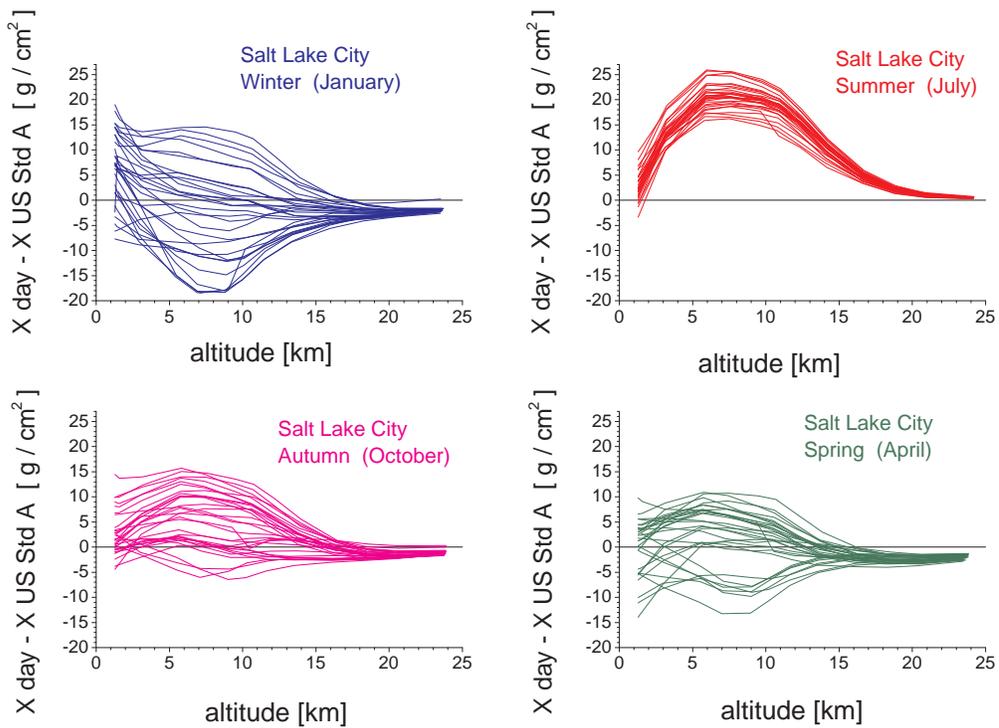}
\end{center}
\caption{Daily profiles of atmospheric depth relative to US Standard Atmosphere in four months of 2002 at Salt Lake City. }
\label{season-daySLC}
\end{figure}

\begin{figure}
\begin{center}
\hspace*{-10mm}
\includegraphics*[width=1.1\textwidth,angle=0,clip]{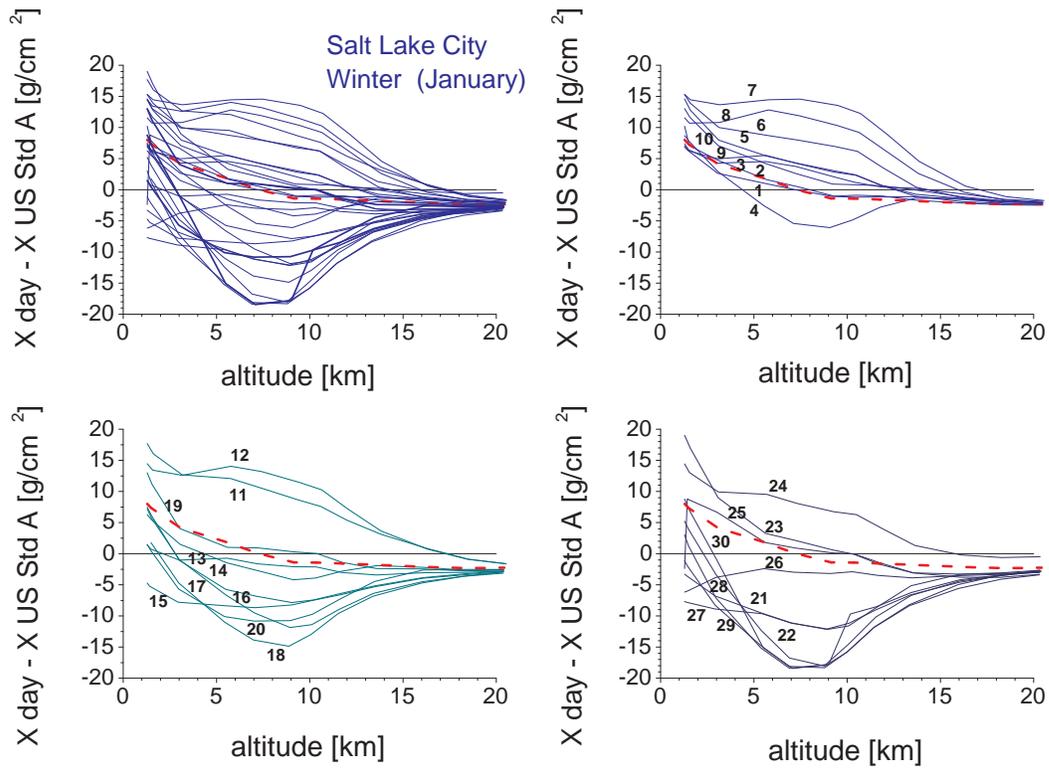}
\end{center}
\caption{Daily profiles of atmospheric depth relative to US Standard Atmosphere for January 2002 at Salt Lake City (upper left). The remaining panels show first, second and third decade of this month separately. The numbers indicate dates of individual profiles. The heavy dashed line represents the monthly average. }
\label{SLC10days}
\end{figure}

In addition to the seasonal variation of the atmosphere, a strong variation is observed on the time scale of days. In Fig. \ref{season-daySLC} the differences of the $X$ distributions relative to the US Standard Atmosphere Model are presented for four seasons in 2002 at Salt Lake City. Atmospheric profiles of individual days are shown for one month in each season. The range of profile variability changes with season: it is largest in winter and smallest in summer. 
In order to examine closer this variability, the individual days of January 2002 are shown in Fig. \ref{SLC10days}. The numbers near the lines denote the date of each day. The four-year average of January is shown by the heavy dashed line. One can note that the profiles of consecutive days differ typically by 3--5 g/cm$^2$, with profiles of consecutive days grouped together. Occasionally, a large change is observed from day to day, with differences exceeding 15 g/cm$^2$ (e.g. days 12--13, 18--19, 22--23). Most of days in the first decade of January lay above the monthly average, while in the rest of the month the profiles of days lay mostly below the monthly average. The deviation of a daily profile from the monthly average can also exceed 15 g/cm$^2$.

\begin{figure}
\begin{center}
\hspace*{-10mm}
\includegraphics*[width=1.1\textwidth,angle=0,clip]{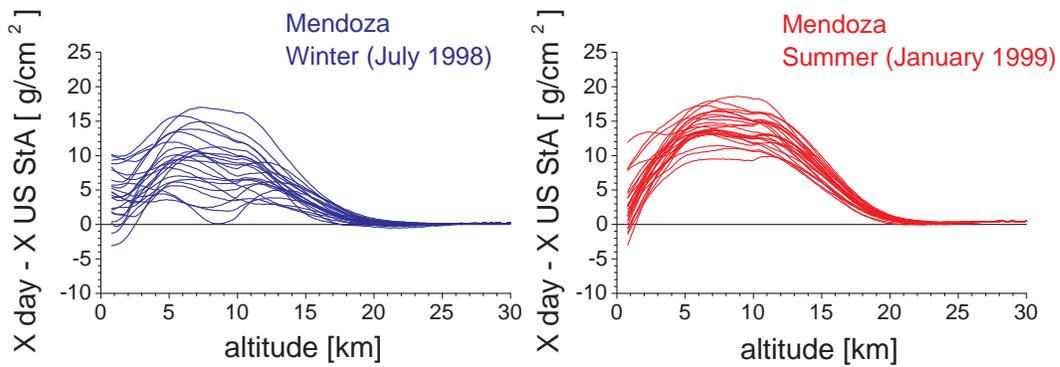}
\caption{\label {Mza-winsum} Daily variation of atmospheric profiles in winter and summer at Mendoza.}
\end{center}
\end{figure}

The character of daily variability in other seasons is very similar to that in winter, but the range of variability is smaller, as seen in Fig. \ref{season-daySLC}. Therefore, the profiles of individual days in seasons other than winter are not explicitly shown here. The character of variation of daily profiles at other sites is similar to that at Salt Lake City. As an example, the daily profiles at Mendoza in winter and summer are shown in Fig. \ref{Mza-winsum}.

\begin{figure}
\begin{center}
\hspace*{-10mm}
\includegraphics*[width=1.1\textwidth,angle=0,clip]{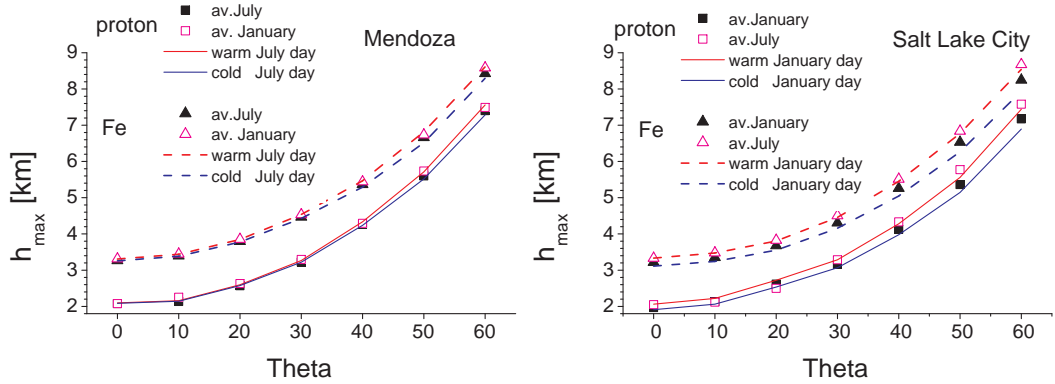}
\caption{\label {shmax} Influence of seasonal and daily variation of atmospheric depth profiles on average altitude of shower maximum of 10 EeV proton  and iron showers as a function of the shower zenith angle. The data points show altitudes of shower maximum in extreme seasons (winter and summer), while the lines represent the extremely warm and cold days in the corresponding single winter month.}
\end{center}
\end{figure}

The data presented in this section demonstrate that taking into account only the seasonal variation of the atmosphere is not sufficient for precise shower reconstruction. The daily variation is equally important and should be accounted for. To illustrate this conclusion, the influence of seasonal and daily variation of atmospheric depth profiles on shower reconstruction is compared in Fig. \ref{shmax}. The altitude of shower maximum is shown for 10 EeV (=10$^{19}$ eV) proton and iron showers, with the $X \leftrightarrow h$ conversion done using the extreme {\em seasonal} profiles (winter-summer) versus the extreme {\em daily} profiles {\em within a single month}. The day-to-day variation of the atmospheric profiles can be seen to be as important as the season-to-season variation. The variation of $h_{max}$ due to fluctuations of daily profiles constitutes a large portion of the $h_{max}$ difference between proton and iron showers, especially at Salt Lake City. 
One can conclude that not only the US Standard Atmosphere model is not sufficiently accurate, but using any {\em average} profile, even made specifically for a particular season at a given site, is not satisfactory. The daily, local soundings to determine the profile of atmospheric depth are recommended for precise determination of the position of shower maximum, and consequently, for identification of cosmic ray primaries.

\section{Day-night variation}

Observations of extensive air showers using the fluorescence technique can be done only during the night. However, some of the meteorological stations usually make balloon soundings during the day only. A question therefore arises whether one can calculate an accurate atmospheric depth profile for the night, based on the radiosonde data collected during the day. The station at Salt Lake City routinely makes two radiosonde soundings during the day (at 11.00 and 12.00 hours) and two soundings during the night (at 23.00 and 0.00 hours). The day-night atmospheric variation can therefore be studied at Salt Lake City. 

\begin{figure}
\begin{center}
\hspace*{-10mm}
\includegraphics*[width=1.1\textwidth,angle=0,clip]{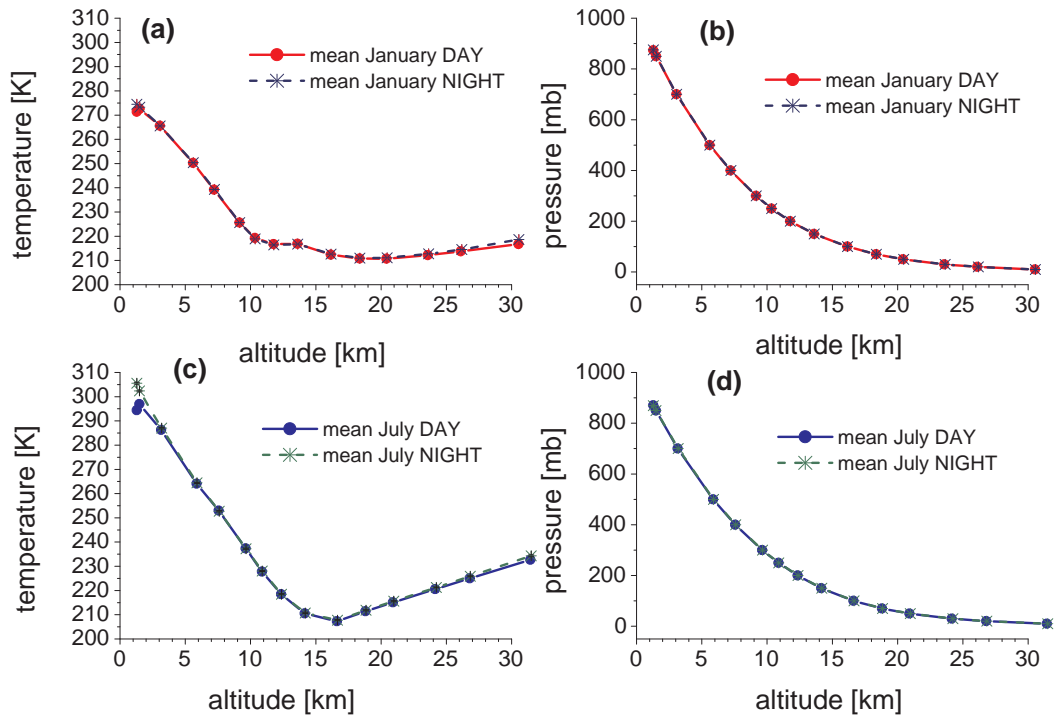}
\end{center}
\caption{Average monthly profiles of temperature and pressure measured during day and during night in January and July at Salt Lake City. }
\label{tp-daynight}
\end{figure}

\begin{figure}
\begin{center}
\includegraphics*[width=0.8\textwidth,angle=0,clip]{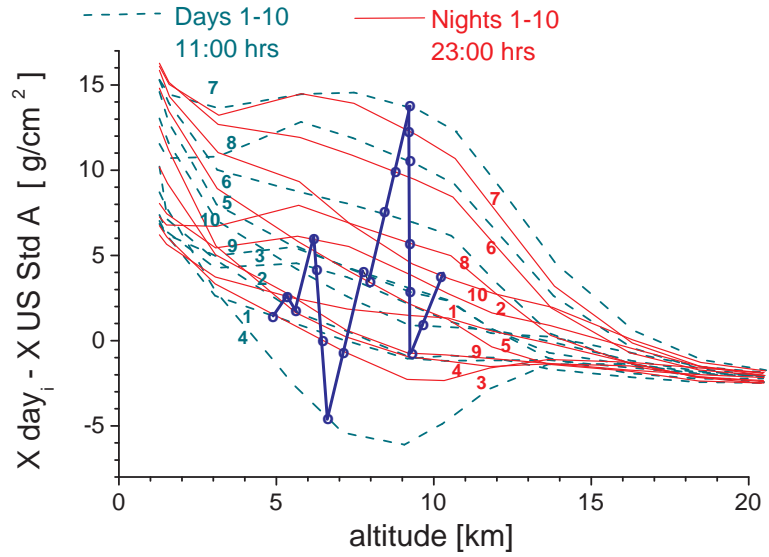}
\end{center}
\caption{Profiles of atmospheric depth measured in the first 10 days (dashed lines) and first 10 nights (thin solid lines) of January 2002 at Salt Lake City. The numbers indicate dates. The heavy solid line shows the sequence of  consecutive day and night profiles.}
\label{daynight10SLC}
\end{figure}

The average (four-year) annual distributions of temperature and pressure for day and night are shown in Fig. \ref{tp-daynight} for January and July at Salt Lake City. The pressure distributions are practically identical, while the temperature shows considerable day-night difference near the ground (the temperature difference at altitudes around 30 km is not as important, as air shower development occurs mostly at lower altitudes). In July this temperature difference near ground is as large as 10 degrees. To demonstrate the day-night differences in profiles of atmospheric depth, the month of January 2002 will be used. Figure \ref{daynight10SLC} shows the atmospheric depth distribution for first ten days of this month and for 10 nights following these days. The plot shows that the profiles of atmospheric depth generally evolve in a continuous way: in most cases a night profile lays between the profiles of the two neighbouring days. Therefore, interpolation between two consecutive days should reproduce the distribution for the night between these days reasonably well.

\begin{figure}
\begin{center}
\includegraphics*[width=0.49\textwidth,angle=0,clip]{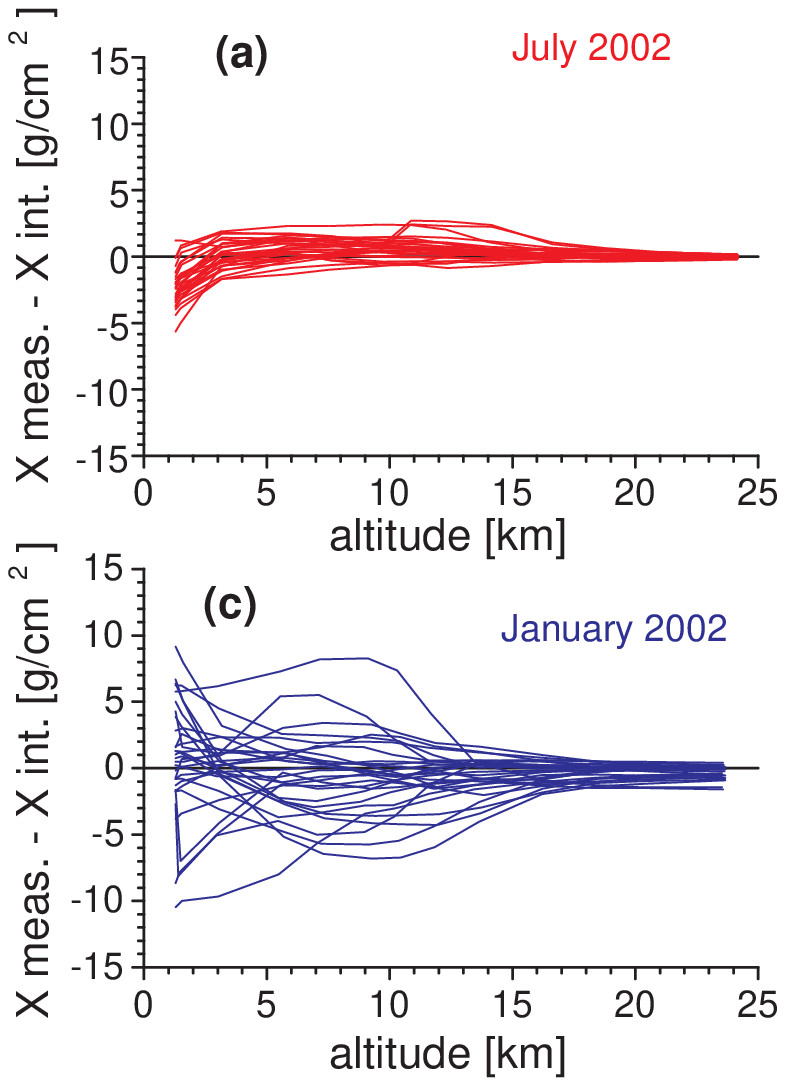}
\hfill
\includegraphics*[width=0.49\textwidth,angle=0,clip]{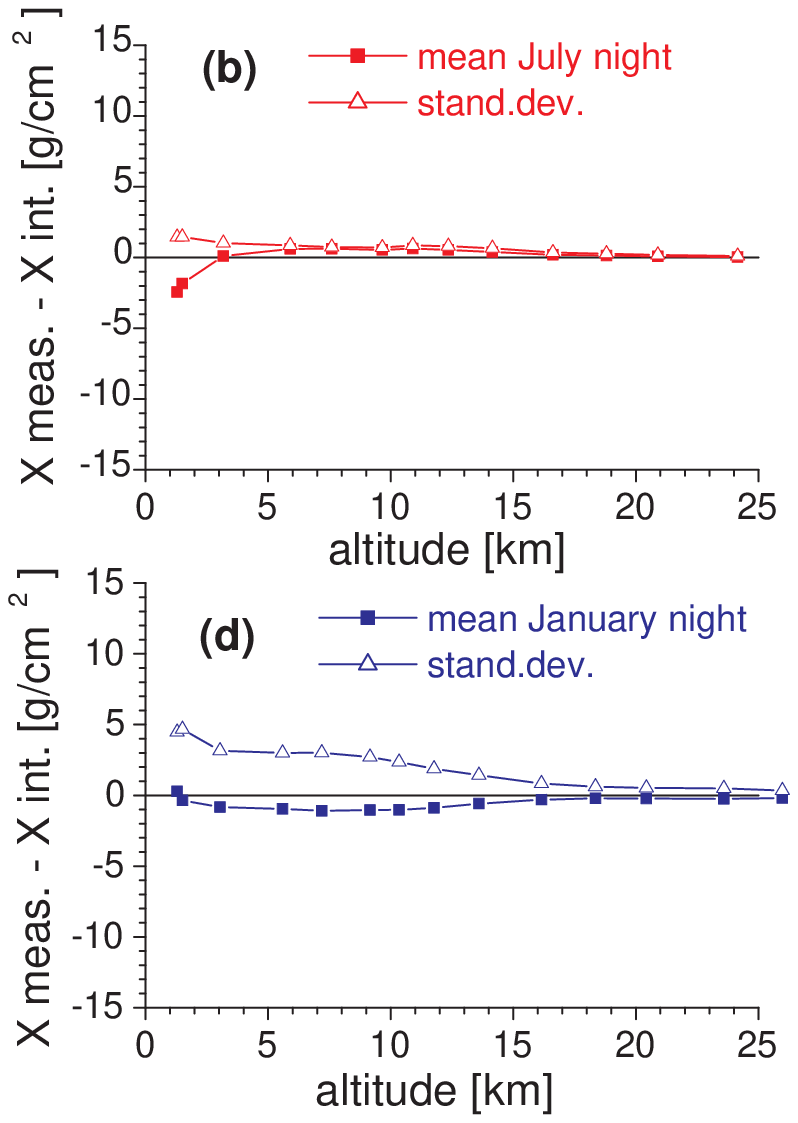}
\end{center}
\caption{{\em (a, c)} Differences between measured and interpolated nigthly profiles of atmospheric depth in summer and winter at Salt Lake City. {\em (b, d)} Mean and standard deviation of profiles shown in {\em (a)} and {\em  (c)}. }
\label{interpolation}
\end{figure}

The comparison of the 'nightly' profiles interpolated from the neighbouring days and those actually measured during the night is shown in Fig. \ref{interpolation} for all nights in January and July 2002 at Salt Lake City. The differences between the interpolated profiles and those actually measured are not large: in most cases they are less than 5 g/cm$^2$. Figures \ref{interpolation}b,d show the mean difference and its standard deviation as a function of altitude. It is evident that only at altitudes below about 2 km above the sea level, i.e. less than 1 km above the ground at Salt Lake City, the interpolated profiles differ appreciably from those actually measured. Therefore, it seems well justified that if soundings of the atmosphere made locally are not available for the night of interest, one can use the soundings made during the neighbouring days. The typical error introduced in this way is less than 3 g/cm$^2$ for altitudes above some 1 km above ground, although for individyal days in winter it can exceed 5 g/cm$^2$.

\section{Summary and conclusion}

The vertical distribution of the atmosphere undergoes large variation at various time scales. Not only the US Standard Atmosphere model, but also any other static model, turn out to be poor approximations of the true state of the atmosphere. The systematic differences in atmospheric depth between the model and the actual value can exceed $X_V=20$ g/cm$^2$. The experimentally determined atmospheric profiles depend on the geographic location, with a different size of seasonal variation at different sites.
The seasonal averages fluctuate from year to year. An average profile for a particular month can vary from year to year by as much as 10 g/cm$^2$. 

The deviation of a particular day's profile from the monthly average can be as large as 15 g/cm$^2$. Although the profiles evolve more or less continuously from day to day, occasionally a large change, on order of 15 g/cm$^2$, occurs between consecutive days. The daily variation of the atmospheric profiles is therefore as important as the seasonal variation, and depends on the geographic location.

The day-night variation appears to be rather small, with the standard deviation of about 3 g/cm$^2$,  and for specific days the difference can exceed 5 g/cm$^2$.

It is to be remembered that the differences in the atmospheric depth quoted above concern the {\em vertical} profiles $X_V(h)$. For real showers, the differences are augmented by the factor $1/\cos \theta$, where $\theta$ is the zenith angle of the shower.

A clear conclusion emerges from these results: using a global model, such as the US Standard Atmosphere, introduces large errors and should be avoided. The daily variation of the atmospheric depth profile is as important as the seasonal variation. For a precise shower reconstruction, the atmospheric depth profile should best be determined locally, near the site of the shower detector, during the night of interest. If the nightly sounding is not available, an interpolation between consecutive days provides a reasonably accurate approximation.
Only when daily profiles are not available, using monthly averages is advisable, with the caveat that the additional uncertainty introduced to shower reconstruction can reach even $\Delta X_V \approx 15$ g/cm$^2$, depending on the season.

\begin{figure}
\begin{center}
\includegraphics*[width=0.7\textwidth,angle=0,clip]{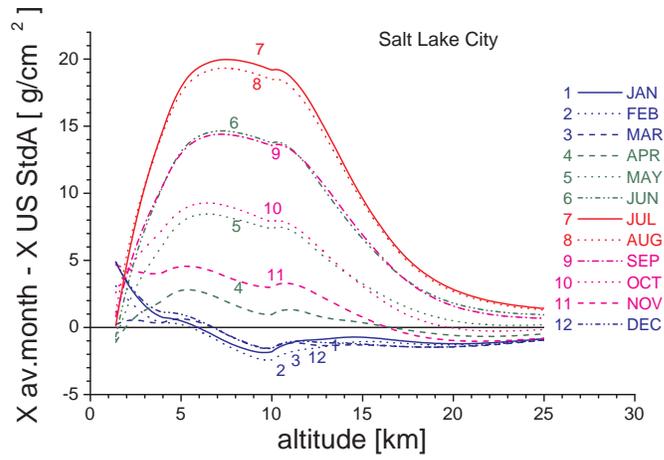}
\includegraphics*[width=0.7\textwidth,angle=0,clip]{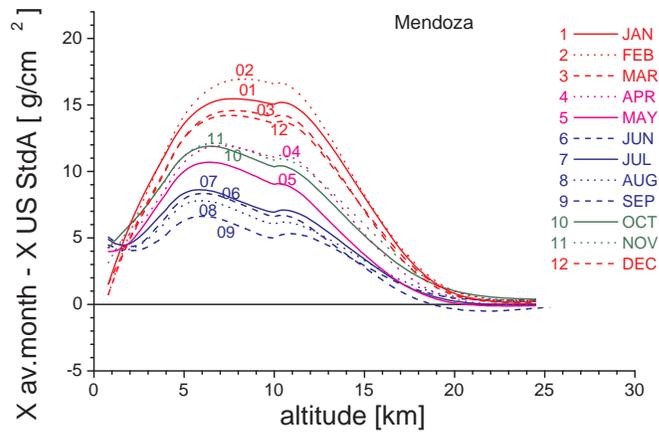}
\end{center}
\caption{Average monthly profiles of atmospheric depth at Salt Lake City and Mendoza. }
\label{seasonsSLCMza}
\end{figure}

If the seasonal averages have to be used, one should note that the traditional division of the year into four 3-month seasons may not be the best choice. As an illustration, in Fig. \ref{seasonsSLCMza} the monthly profiles (4-year averages) are shown for Salt Lake City and Mendoza. At both sites the profiles of four winter months (December--March at Salt Lake City and June--September at Mendoza) are close to each other. Thus from the point of view of shower reconstruction, these four months constitute the winter season. Similarly, the summer months might be July--August at Salt Lake City and December--March at Mendoza. The parameterization of these monthly profiles is given in Tables 1 and 2: the parameters of the exponential functions of Eq. 2 are given in ranges of altitude $h$: 0--7 km, 7--12 km, 12--30 km and 30--100 km. These parameterizations approximate the monthly averages reasonably well, with the typical deviation from the data on the order of 2 g/cm$^2$. However, one should keep in mind that when using the monthly averages, the daily variation of the atmosphere (not accounted for) is the main source of uncertainties.

{\em Acknowledgements.} We thank Hans Klages, Ralph Engel and Bianca Keilhauer for useful discussions. This work was partially supported by the Polish Committee for Scientific Research under grants No. PBZ KBN 054/P03/2001 and 2P03B 11024 and in Germany by the DAAD under grant No. PPP 323. MR is supported by the Alexander von Humboldt Foundation.

\clearpage

\begin{table}
\caption{\label{tab1 }Parameters of fits to Eq.2 for annual and monthly averages of January--June at Mendoza and Salt Lake City.  }

\scriptsize
\begin{center}
\begin{tabular}{|cc|ccc|ccc|}

\hline
\hline
Month &  altitude & \multicolumn{3}{c|}{Mendoza} & \multicolumn{3}{c|}{Salt Lake City}\\
 &  km &$a_i$ (g/cm$^2$) & $b_i$ (g/cm$^2$) & $c_i$ (cm) & $a_i$ (g/cm$^2$) & $b_i$ (g/cm$^2$) & $c_i$ (cm) \\
\hline
 av. year &  0--7 & -135.3657 & 1173.5795 & 964481.07 & -150.5405 & 1186.6019 & 975370.45 \\
 & 7--12 & -71.2233 & 1148.3744 & 849610.95 & -67.6044 & 1149.0158 & 833226.65 \\
  & 12--30 & 1.2055 & 1429.8293 & 621337.57 & 0.9984 & 1388.8984 & 624933.79 \\
  & 30--100 & 0.0106 & 870.8015 & 708709.48 & 0.0107 & 872.0169 & 705528.85 \\
\hline
January  	 &	  0--7 &	-143.1356&	1179.0132&	986940.77&	-124.0319&	1170.7336&	913948.11 \\
&	 7--12 &	-73.9046&	1144.1639&	868109.44&	-52.5789&	1150.1929&	787455.67 \\
&	12--30 &	1.2702&	1479.2616&	617703.59&	0.5428&	1319.9111&	631652.82 \\
&	30--100&	0.0105&	806.3091&	723598.11&	0.0108&	887.3348&	696567.06\\
\hline

February 	&  0--7 &	-135.9059&	1170.4334&	982287.73&	-132.5538&	1176.2907&	926334.89 \\
&	 7--12& 	-83.9765&	1144.0362&	892118.25&	-53.1326&	1152.4537&	786177.03 \\
&	12--30& 	1.4610&	1497.6338&	614675.42&	0.4332&	1311.1860&	632894.01 \\
&	30--100&	0.0105&	809.6089&	723681.98&	0.0108&	876.8543&	697419.09 \\
\hline
March   	&  0--7 &	-133.5557&	1169.3232&	972625.45&	-145.2204&	1183.7779&	948368.62 \\
&	 7--12& 	-77.3797&	1144.7436&	872448.79&	-59.3777&	1151.4365&	800062.39 \\
&	12-30& 	1.4111&	1477.8039&	615783.26&	0.7382&	1331.2019&	627787.50 \\
&	30--100&	0.0106&	851.9467&	713538.03&	0.0107&	829.3134&	707135.34 \\
\hline
April    &  0--7& 	-124.4756&	1164.0947&	950839.84&	-158.6882&	1193.1643&	974561.42 \\
&	 7--12& 	-73.4197&	1147.2795&	856075.22&	-60.4142&	1150.4190&	807551.59 \\
&	12--30& 	1.3443&	1448.0295&	617898.74&	0.7418&	1337.5534&	630095.46 \\
&	30--100&	0.0107&	896.5384&	703568.67&	0.0106&	802.2364&	716323.66 \\
\hline
 May     &	  0--7& 	-136.2964&	1174.7211&	962756.90& -163.3301&	1194.5484&	998423.61 \\
&	 7--12& 	-75.1491&	1152.7025&	851469.53&	-73.3103&	1148.9595&	846457.10 \\

&	12--30& 	1.2284&	1426.3171&	618974.94&	1.2485&	1393.7719&	624157.30 \\
&	30--100&	0.0107&	918.3066&	697152.59&	0.0105&	803.1666&	722517.74 \\
\hline
June    &	0--7& 	-128.8113&	1169.5907&	945622.84&	-172.3404&	1200.2392&	1028838.67 \\
&	 7--12 &	-68.7814&	1151.9990&	834385.28&	-82.4338&	1149.3884&	878122.45 \\
&	12--30& 	1.2704&	1409.3793&	619897.12&	1.5904&	1443.0171&	621147.73 \\
&	30--100&	0.0108&	961.9427&	689571.21&	0.0104&	817.2777&	725898.04 \\
\hline
\hline
\end{tabular}
\end{center}
\normalsize

\end{table}

\begin{table}
\caption{\label{tab2 }Parameters of fits to Eq.2 for  monthly averages of July--December at Mendoza and Salt Lake City. }

\scriptsize
\begin{center}
\begin{tabular}{|cc|ccc|ccc|}

\hline
\hline
Month &  altitude & \multicolumn{3}{c|}{Mendoza} & \multicolumn{3}{c|}{Salt Lake City}\\
 &  km &$a_i$ (g/cm$^2$) & $b_i$ (g/cm$^2$) & $c_i$ (cm) & $a_i$ (g/cm$^2$) & $b_i$ (g/cm$^2$) & $c_i$ (cm) \\
\hline
July     &	  0--7& 	-137.0555&	1177.1753&	956627.92&	-186.6534&	1212.2804&	1063017.00 \\
&	 7--12& 	-64.9589&	1151.1233&	827880.59&	-84.8974&	1149.8003&	894551.28 \\
&	12--30& 	1.0641&	1395.2895&	624326.08&	1.6325&	1487.4886&	619070.90 \\
&	30--100&	0.0108&	944.3901&	693374.71&	0.0105&	865.8201&	718581.76 \\
\hline
August   &	  0--7& 	-134.8719&	1174.8045&	952297.63&	-180.1574&	1207.9734&	1050566.85 \\
&	 7--12& 	-63.5567&	1151.4517&	823187.06&	-83.3827&	1150.0202&	889764.14 \\
&	12--30& 	0.9910&	1371.1796&	628951.14&	1.4645&	1478.0912&	620467.75 \\
&	30--100&	0.0107&	921.2122&	699108.09&	0.0106&	892.6000&	711914.74 \\
\hline
September &	  0--7& 	-130.0794&	1171.1373&	942652.20&	-163.0811&	1193.8657&	1013663.92 \\
&	 7--12& 	-60.0068&	1150.2624&	814932.30&	-77.4142&	1148.1548&	868479.61 \\
&	12---30& 	1.1616&	1368.4099&	628284.06&	1.1586&	1439.6748&	623016.99 \\
&	30--100&	0.0107&	903.4670&	703442.29&	0.0107&	898.7723&	705833.25 \\
\hline
October   &	  0--7& 	-138.3774&	1177.5028&
	967536.19&	-146.6575&	1183.5344&	973580.38 \\
&	 7--12& 	-71.6140&	1151.7126&	847818.42&	-74.4407&	1150.8957&	848663.14 \\
&	12--30& 	1.1403&	1414.0158&	624388.32&	1.1837&	1417.8501&	619323.37 \\
&	30--100&	0.0106&	870.0336&	709894.94&	0.0108&	921.7123&	695498.81 \\
\hline
November &	  0--7& 	-140.1651&	1178.5255&	971957.76&	-131.1912&	1174.5426&	936106.26 \\
&	 7--12& 	-69.6025&	1149.7162&	846709.27&	-63.3954&	1149.1226&	817832.83 \\
&	12--30& 	0.8541&	1412.1573&	626853.42&	0.6643&	1377.7564&	623196.00 \\
&	30--100&	0.0105&	809.6100&	721090.69&	0.0109&	983.9849&	678655.75 \\
\hline
December &	  0--7& 	-143.6484&	1178.6584&	985389.83&	-125.1286&	1171.2049&	916604.22 \\
&	 7--12& 	-75.3706&	1145.6133&	867024.95&	-56.5295&	1151.9447&	794149.64 \\
&	12--30& 	1.2267&	1460.0256&	619529.43&	0.5446&	1330.9703&	628346.13 \\
&	30--100&	0.0104&	790.1251&	726748.15&	0.0108&	932.9255&	686075.79 \\
\hline
\hline
\end{tabular}
\end{center}
\normalsize

\end{table}

\end{document}